\documentclass[epjCONF]{svjour}
\usepackage{graphics}
\usepackage[varg]{txfonts} 
\usepackage[latin1]{inputenc}
\session-title{Progress on Neutron-Target Multipoles above 1~GeV}
\begin{document}
\title{Progress on Neutron-Target Multipoles above 1~GeV}
\author{I.I.~Strakovsky\inst{1}\fnmsep\thanks{\email{igor@gwu.edu}} 
\and W.~Chen\inst{2} 
\and H.~Gao\inst{2} 
\and W.J.~Briscoe\inst{1}
\and D.~Dutta\inst{3} 
\and A.E.~Kudryavtsev\inst{4}
\and M.~Mirazite\inst{5}
\and M.~Paris\inst{6}
\and P.~Rossi\inst{5}
\and S.~Stepanyan\inst{7}
\and V.E.~Tarasov\inst{4}
\and R.L.~Workman\inst{1}}
\institute{The George Washington University, Washington, DC 20052, USA
\and Duke University, Durham, NC 27708, USA
\and Mississippi State University, Mississippi State, MS 39762, USA
\and Institute of Theoretical and Experimental Physics, Moscow, 117259 Russia
\and INFN, Laboratory Nazionali di Frascati, 00044 Frascati, Italy
\and Theory Division, Los Alamos National Laboratory, Los Alamos, NM 87545, USA
\and Thomas Jefferson National Accelerator Facility, Newport News, VA 23606, USA}
\abstract{
We report a new extraction of nucleon resonance couplings using 
$\pi^-$ photoproduction cross sections on the neutron. The world 
database for the process $\gamma n\to\pi^-p$ above 1~GeV has 
quadrupled with the addition of new differential cross sections
from the CEBAF Large Acceptance Spectrometer (CLAS) at Jefferson Lab 
in Hall~B.  Differential cross sections from CLAS have been improved 
with a new final-state interaction determination using a diagrammatic
technique taking into account the SAID phenomenological $NN$ and 
$\pi N$ final-state interaction amplitudes.  Resonance couplings have 
been extracted and compared to previous determinations.  With the 
addition of these new cross sections, significant changes are seen 
in the high-energy behavior of the SAID cross sections and 
amplitudes.
}
\maketitle

The GW group is focusing on the study of $NN$ and $\pi N$ elastic 
scattering as well as on the study of $\gamma N\to\pi (\eta)N$ reactions.  
The former subject was devoted to study of electro-magnetic (EM) couplings 
$N^\ast (\Delta^\ast)\to\gamma N$.  The radiative decay width of the 
neutral $N^\ast$- and $\Delta^\ast$-states may be extracted from $\pi^-$ 
and $\pi^0$ photoproduction on the neutron target (typically the deuteron) 
and requires the use of the model dependent nuclear corrections 
[final-state interaction (FSI)].  As a result, our knowledge of the 
neutral resonance decays is less precise compared to the charged ones.

For today, the experimental information on the reactions on the proton
is more reach then that on the neutron (15\%)~\cite{SAID}.  Only with
good data on both proton and neutron targets, one can hope to disentangle
the isoscalar and isovector EM couplings of various $N^\ast$ and 
$\Delta^\ast$ resonances~\cite{old}, as well as the isospin properties
of the non-resonant background amplitudes.

Partially, it is compensated by experiments on pionic beams, e.g., 
$\pi^-p\to\gamma n$ (not $\pi^0n\to\gamma n$) as Crystal Ball 
Collaboration made, for instance, at BNL~\cite{aziz} for the inverse
photon energy $E_\gamma$=285 -- 689~MeV and $\theta$=41 -- 
148$^\circ$.  This process is free from complications associated with 
the deuteron target.  However, the disadvantage of using the reaction
$\pi^-p\to\gamma n$ is the 5 to 500 times larger cross sections for
$\pi^-p\to\pi^0n\to\gamma\gamma n$ depending on pion kinetic energy
$T_\pi$ and $\gamma$ production angle $\theta$.  That is why, we are 
forcing to use the deuteron as the neutron target.

\begin{figure}\sidecaption
\resizebox{0.16\hsize}{!}{\includegraphics*{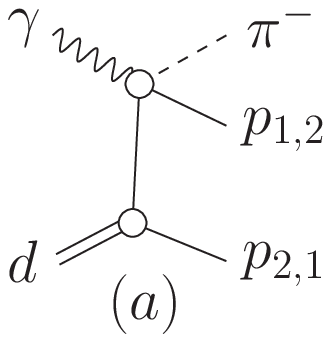}~~~~~}
\resizebox{0.18\hsize}{!}{\includegraphics*{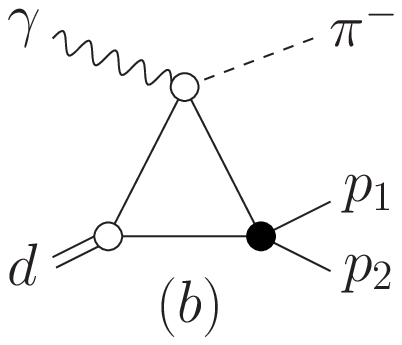}~~~~~}
\resizebox{0.18\hsize}{!}{\includegraphics*{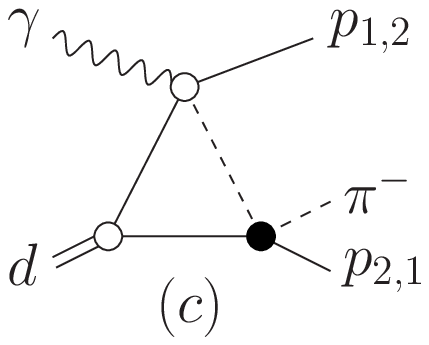}}
\caption{Feynman diagrams for the leading components of the
        $\gamma d\to\pi^-pp$ amplitude. (a) IA, (b) $pp$-FSI, 
	and (c) $\pi$N-FSI. Filled black circles show FSI
	vertices. Wavy, dashed, solid, and double lines 
	correspond to the photons, pions, nucleons, and 
	deuterons, respectively.} \label{fig:1}
\end{figure}

In a further study of the FSI corrections for the $\gamma n\to\pi^-p$
cross section determination from the deuteron data, we used a 
diagrammatic technique~\cite{FSI}.  The Feynman diagrams corresponding 
to the Impulse Approximation (IA) [Fig.~\ref{fig:1}(a)] and $pp$-FSI 
[Fig.~\ref{fig:1}(b)], and $\pi$N-FSI [Fig.~\ref{fig:1}(c)] amplitudes
for the reaction $\gamma d\to\pi^-pp$ is shown in Fig.~\ref{fig:1}.
IA and $\pi$N-FSI diagrams [Figs.~\ref{fig:1}(a),(c)] include also 
the cross-terms  
final protons. 

The $\gamma N\!\to\!\pi N$ amplitudes were expressed through four 
independent Chew-Goldberger-Low-Nambu (CGLN) amplitudes~\cite{CGLN}, 
which were generated by the SAID code, using the GW DAC pion 
photoproduction multipoles~\cite{pr_PWA}. The $N\!N$- and 
$\pi N$-scattering amplitudes were calculated, using the results of 
GW $N\!N$~\cite{NN} and $\pi N$~\cite{piN} PWAs. The DWF was taken 
from the Bonn potential (full model) \cite{Bonn}.  The elementary 
amplitudes are dependent on the momenta of the external and 
intermediate particles in Fig.~\ref{fig:1}. Thus, Fermi motion is 
taken into account in the $\gamma d\to\pi^-pp$ amplitude.

We applied FSI corrections~\cite{FSI} dependent on the E$_\gamma$
and pion production $\theta$ and taking into account a kinematical 
cut with momenta less (more) than 200~MeV/$c$ for slow (fast) 
outgoing protons.  Overall, the FSI correction factor $R< 1$, while 
the effect, i.e., the $(1-R)$ value, is less than 10\% and the 
behavior is very smooth vs. pion production angle.  

The contribution of FSI calculations~\cite{FSI} to the overall
systematics is estimated to be 2\% (3\%) below (above) 1800~MeV.
Then we added FSI systematics to the overall experimental systematics 
in quadrature.

\begin{figure}\sidecaption
\resizebox*{!}{7cm}{%
  \rotatebox{90}{%
    \includegraphics{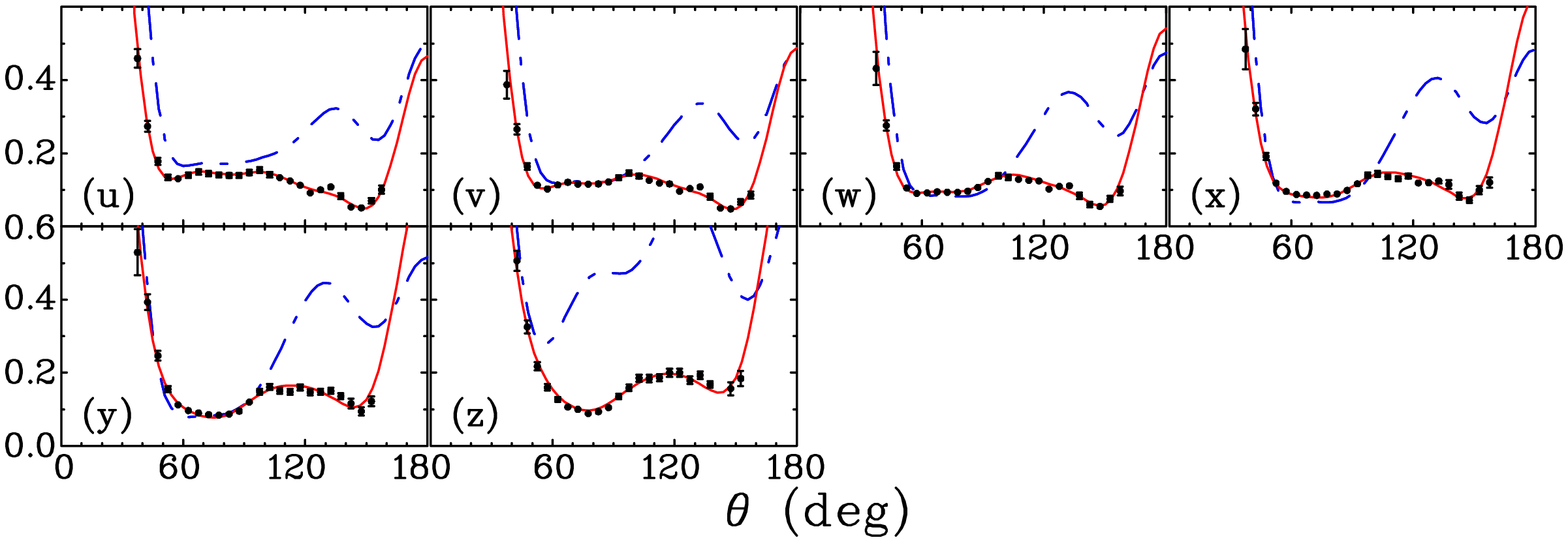}\hfill%
    \includegraphics{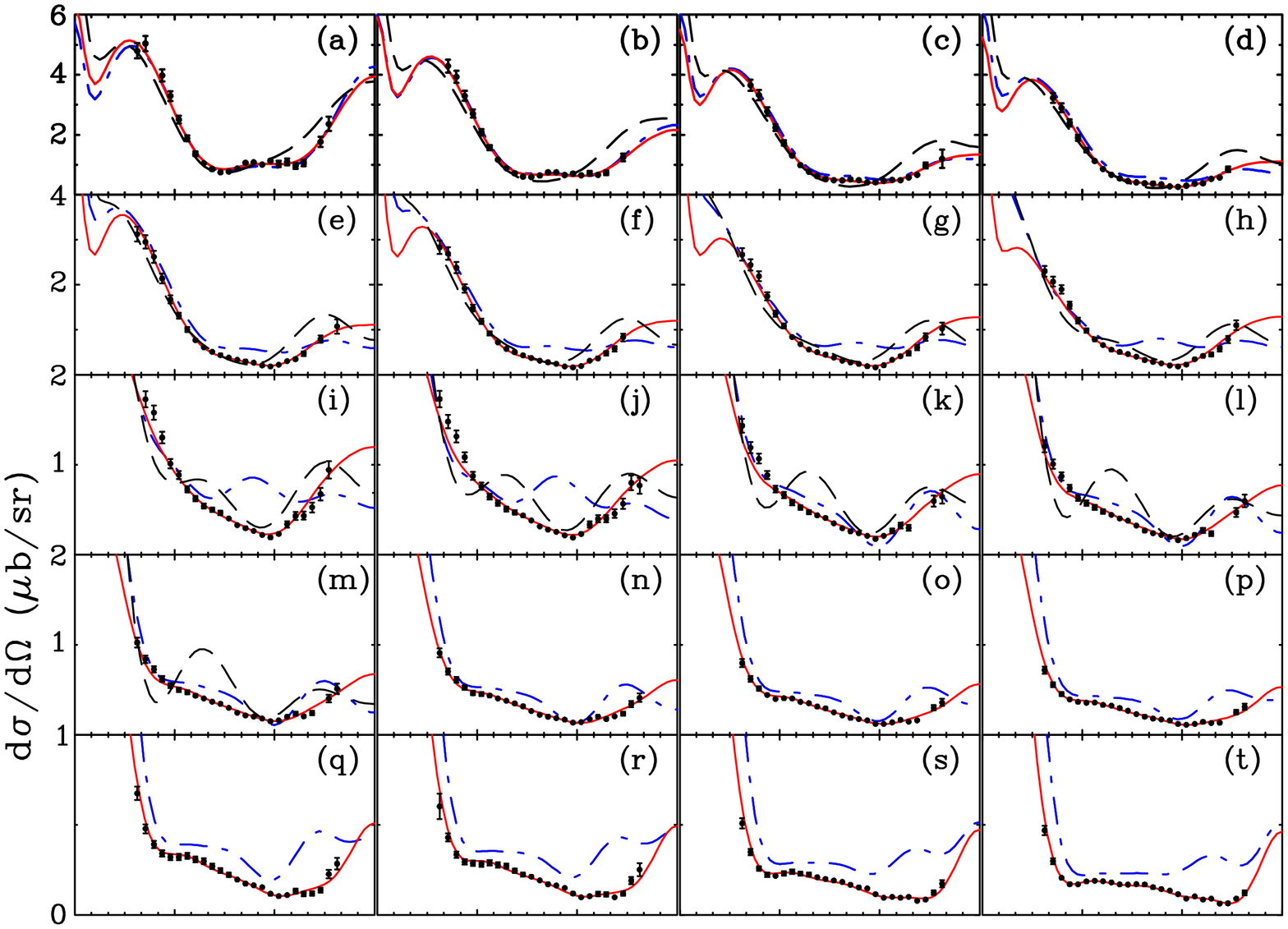}
  }%
}
\caption{The differential cross section for $\gamma n\to\pi^-p$
        below E$_\gamma$ = 2.7~GeV versus pion CM angle.
        Solid (dash-dotted) lines correspond to the GB12
        (SN11~\protect\cite{sn11}) solution. Dashed lines
        give the MAID07~\protect\cite{maid} predictions.
        Experimental data are from the current (filled
        circles). Plotted uncertainties are statistical.
        (a) E = 1050~MeV,        (b) E = 1100~MeV,
        (c) E = 1150~MeV,        (d) E = 1200~MeV,
        (e) E = 1250~MeV,        (f) E = 1300~MeV,
        (g) E = 1350~MeV,        (h) E = 1400~MeV,
        (i) E = 1450~MeV,        (j) E = 1500~MeV,
        (k) E = 1550~MeV,        (l) E = 1600~MeV,
        (m) E = 1650~MeV,        (n) E = 1700~MeV,
        (o) E = 1750~MeV,        (p) E = 1800~MeV,
        (q) E = 1850~MeV,        (r) E = 1900~MeV,
        (s) E = 2000~MeV,        (t) E = 2100~MeV,
        (u) E = 2200~MeV,        (v) E = 2300~MeV,
        (w) E = 2400~MeV,        (x) E = 2500~MeV,
        (y) E = 2600~MeV, and    (z) E = 2700~MeV.}
	\label{fig:2}
\end{figure}
We have included the new cross sections from the CLAS experiment
\cite{CLAS} in a number of multipole analyses covering incident
photon energies up to 2.7~GeV, using the full SAID database, in
order to gauge the influence of these measurements, as well as their
compatibility with previous measurements~\cite{gb12}.

A comparison of the CLAS data with fits and predictions is given in
Fig.~\ref{fig:2}. It is interesting to note that the data appear to
have fewer angular structures than the earlier fits. The overall
$\chi^2$ has remained stable against the growing database, which
has increased by a factor of 2 since 1995 (most of this increase
comes from data from photon-tagging facilities).

In fitting the data, the stated experimental systematic
uncertainties have been used as an overall normalization adjustment
factor for the angular distributions~\cite{sn11}. Presently, the
pion photoproduction database below E$_\gamma$ = 2.7~GeV consists
of 26179 data points that have been fit in the GB12 (GZ12~\cite{mc12}) 
solution with $\chi^2$ = 54832 (50998).  The contribution to the 
total $\chi^2$ in the GB12 (GZ12) analyses of the $626$ new CLAS
$\gamma n\to\pi^-p$ data points (e.g., those data points up to
E$_\gamma$ = 2.7~GeV) is 1580 (1190).  This should be compared to
a starting $\chi^2$ = 45636 for the new CLAS data using predictions
from our previous SN11 solution~\cite{sn11}.

\begin{figure}\sidecaption
\resizebox*{!}{2.3cm}{%
  \rotatebox{90}{%
    \includegraphics{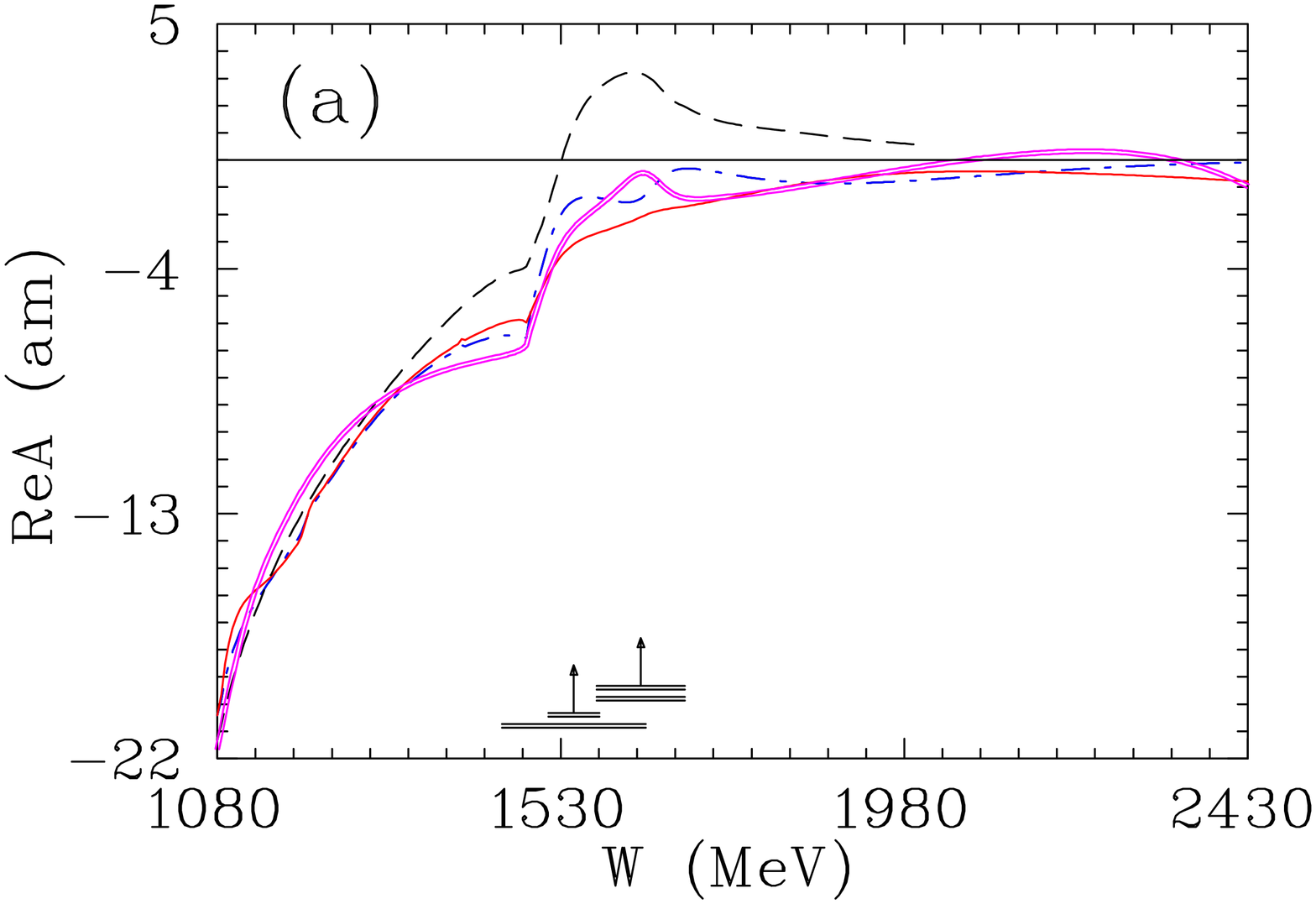}%
  }%
}
\resizebox*{!}{2.3cm}{%
  \rotatebox{90}{%
    \includegraphics{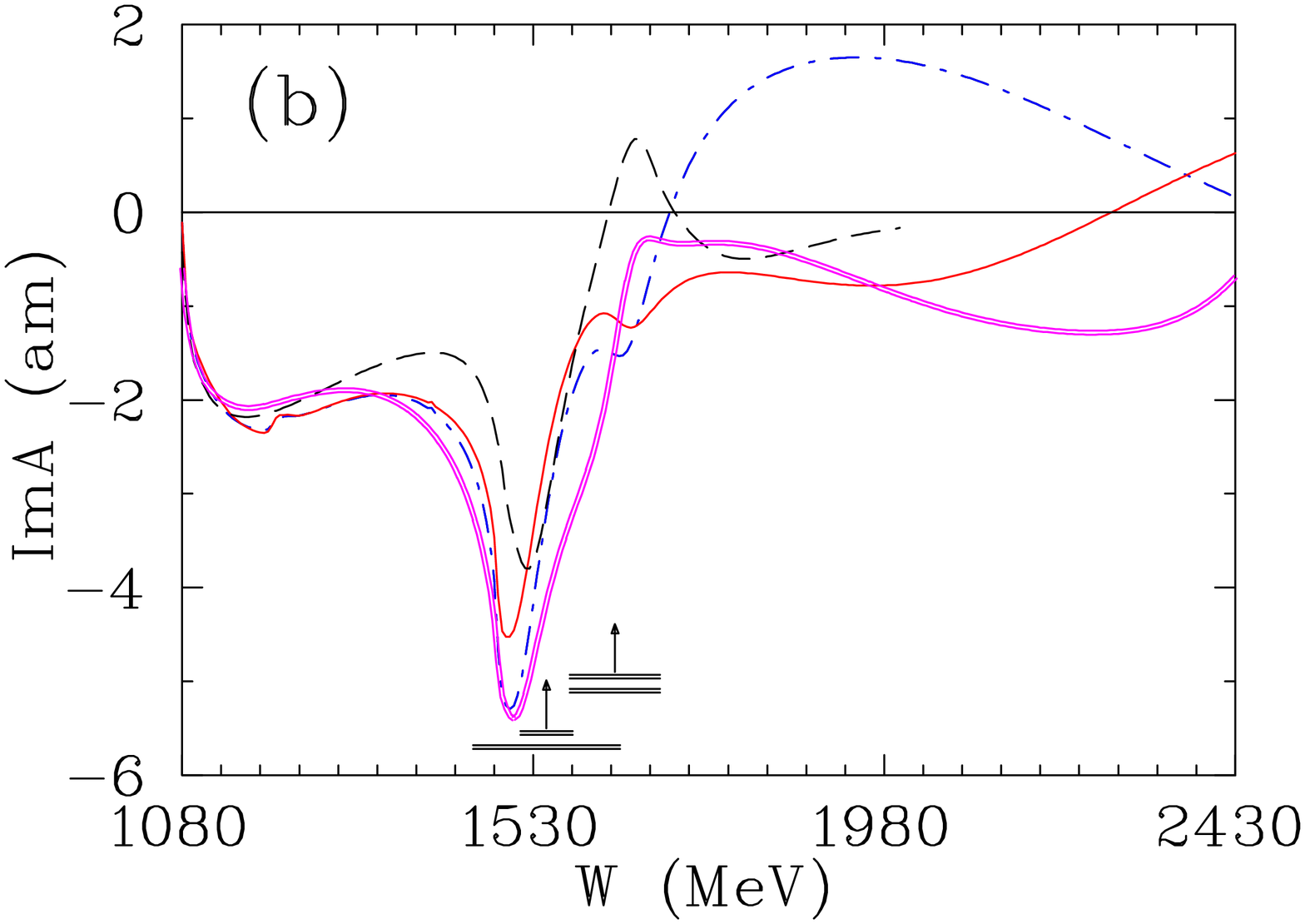}%
  }%
}
\resizebox*{!}{2.3cm}{%
  \rotatebox{90}{%
    \includegraphics{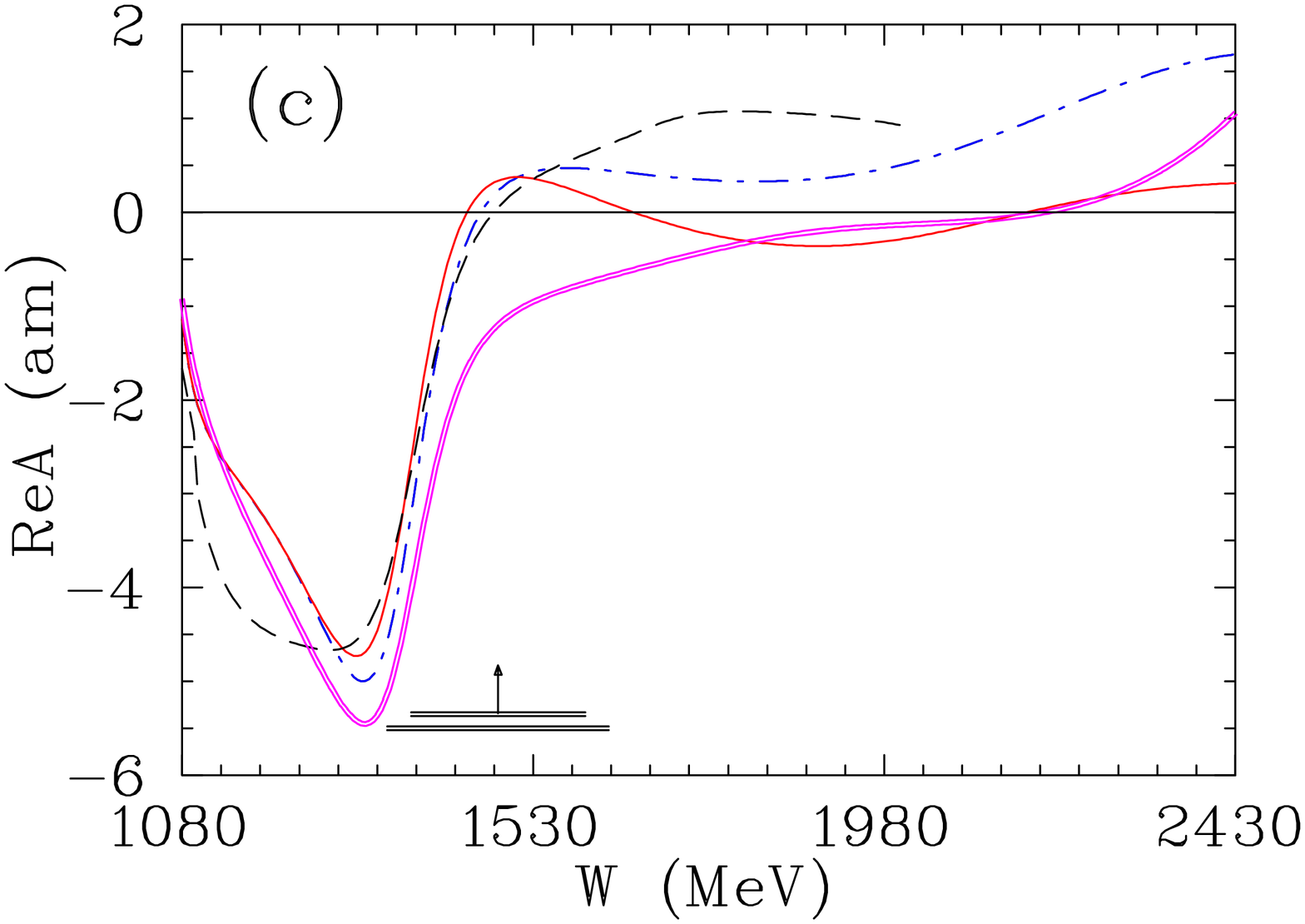}%
  }%
}
\resizebox*{!}{2.3cm}{%
  \rotatebox{90}{%
    \includegraphics{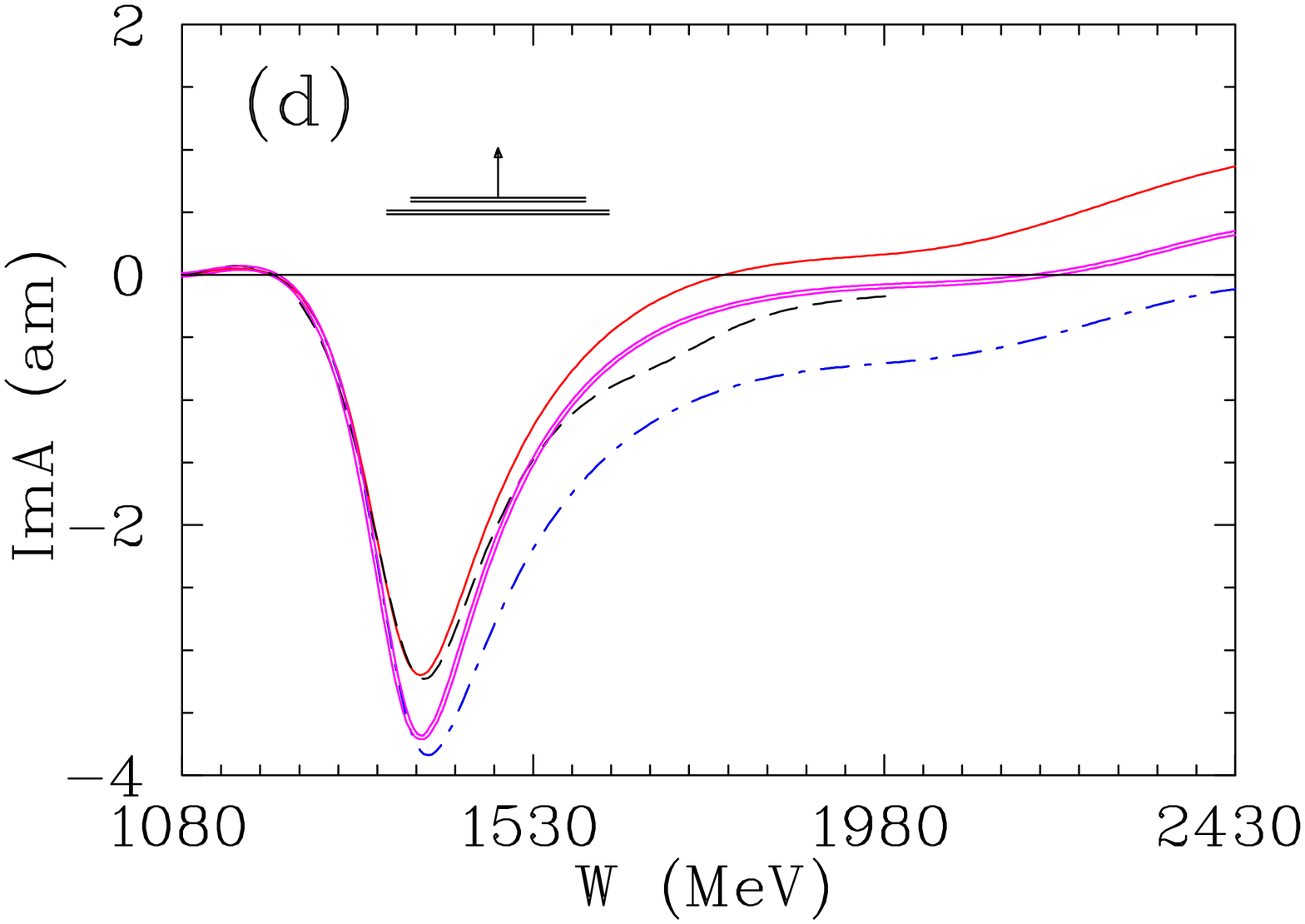}%
  }%
}
\resizebox*{!}{2.4cm}{%
  \rotatebox{90}{%
    \includegraphics{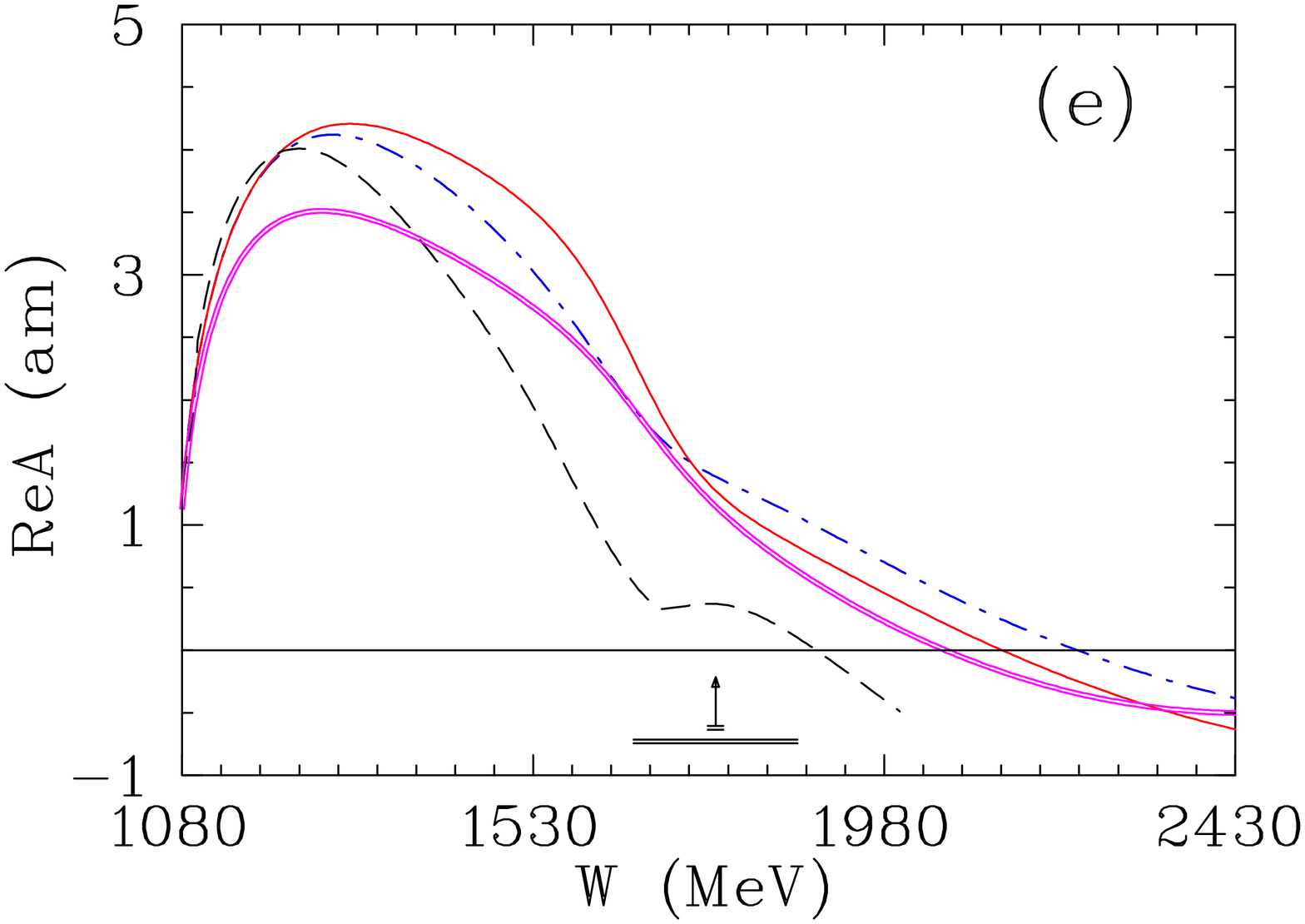}%
  }%
}
\resizebox*{!}{2.4cm}{%
  \rotatebox{90}{%
    \includegraphics{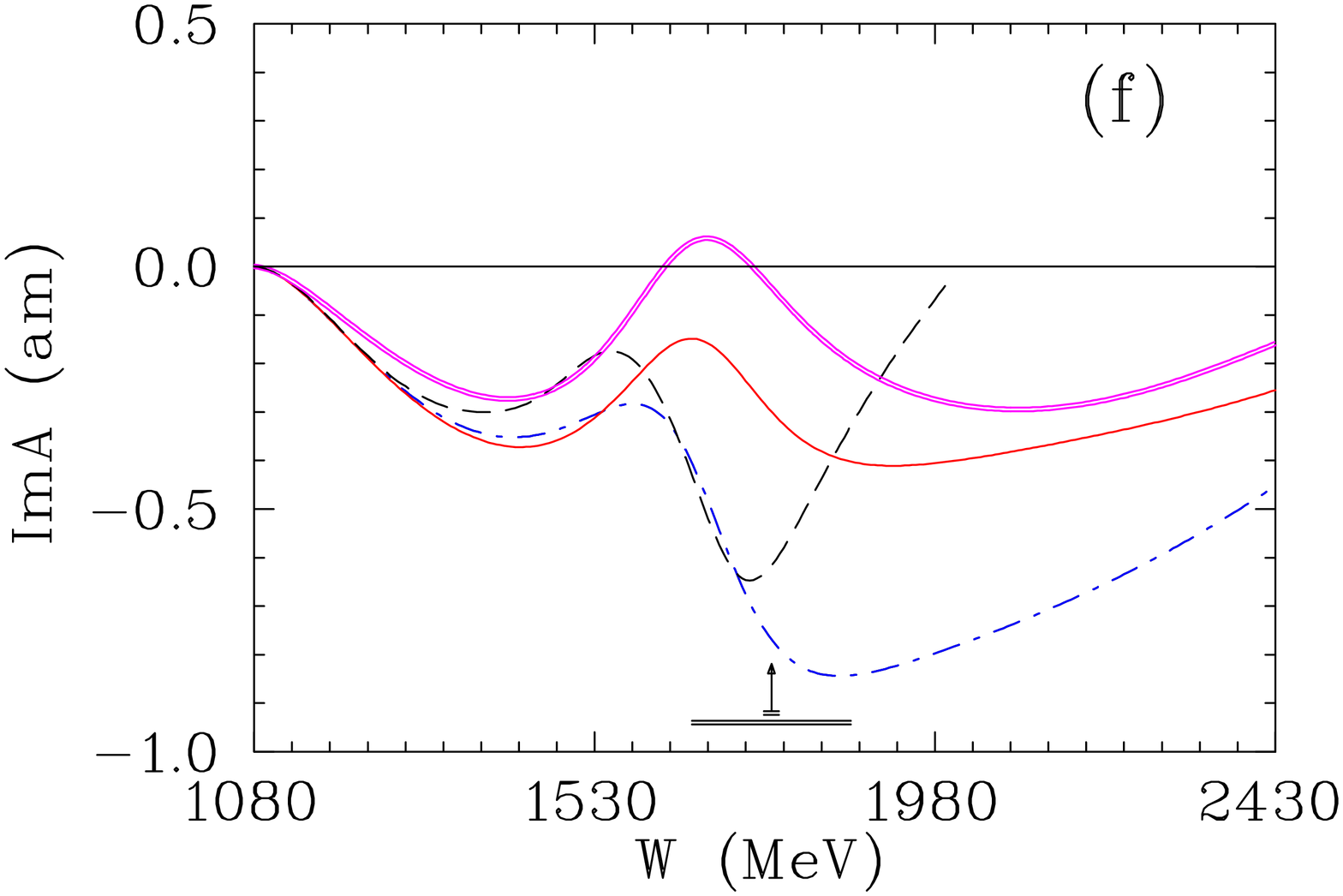}%
  }%
}
\resizebox*{!}{2.4cm}{%
  \rotatebox{90}{%
    \includegraphics{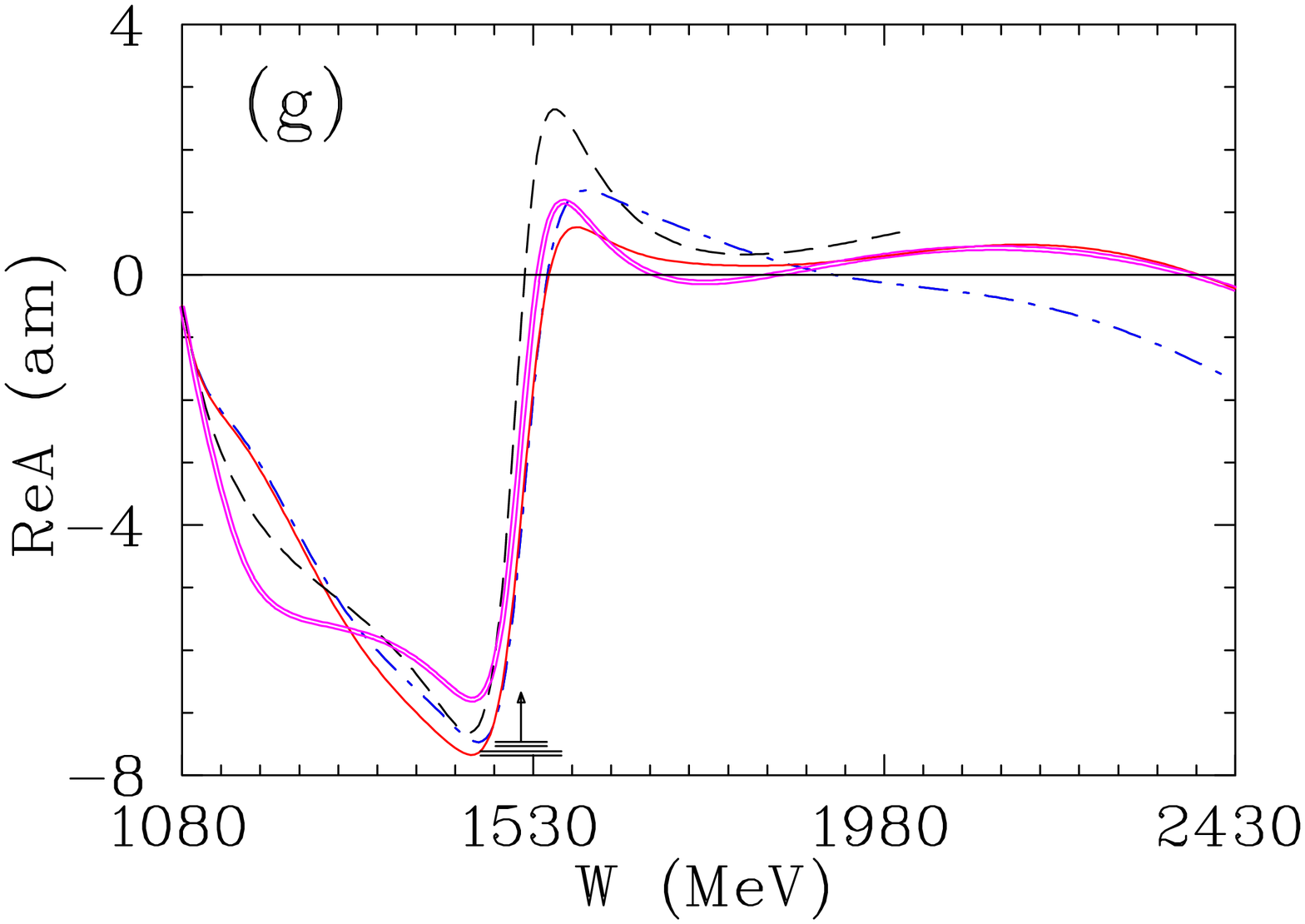}%
  }%
}
\resizebox*{!}{2.4cm}{%
  \rotatebox{90}{%
    \includegraphics{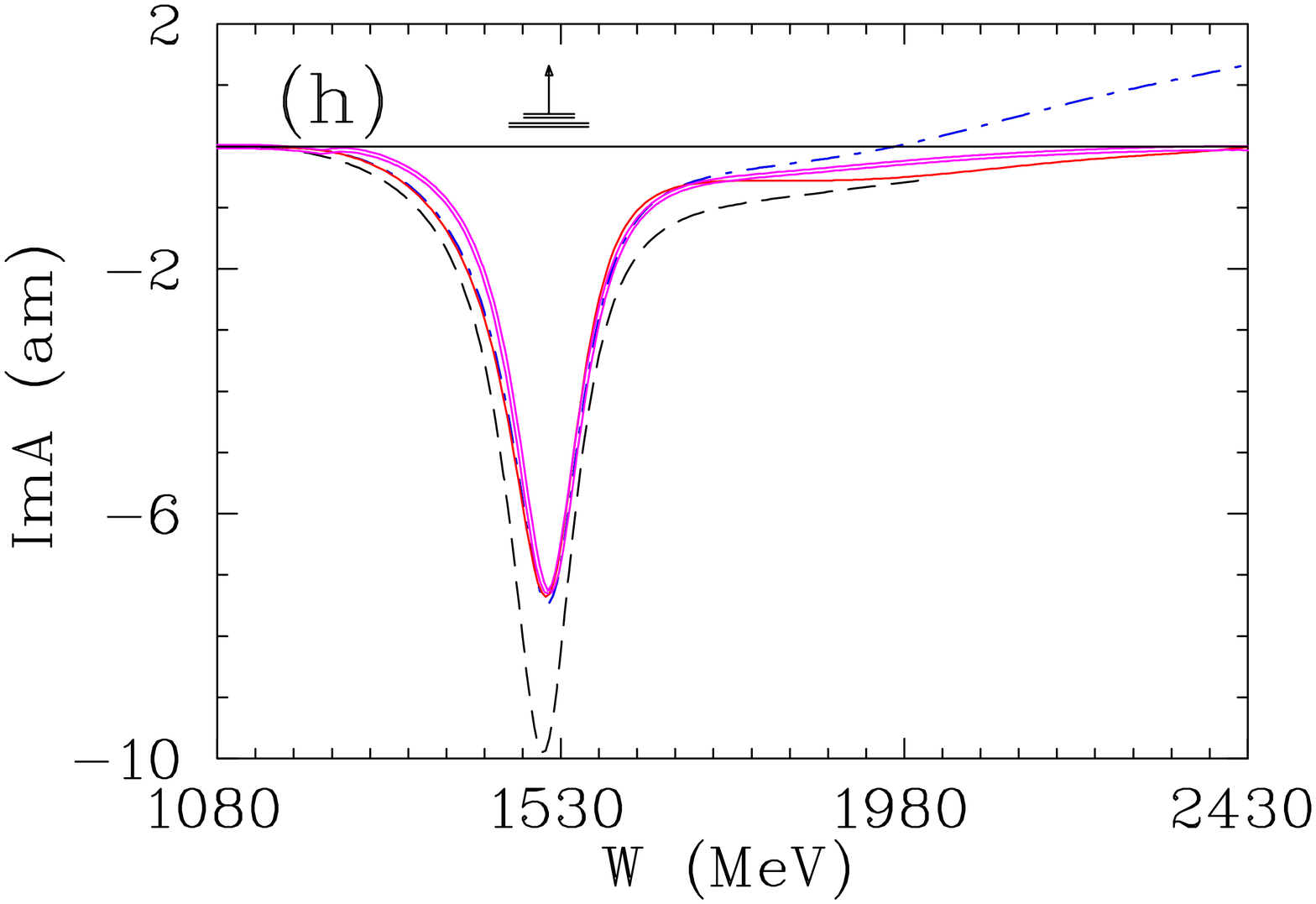}%
  }%
}
\caption{Dominant neutron multipole I=1/2 amplitudes from
        threshold to W = 2.43~GeV (E$_{\gamma}$ = 2.7~GeV).
        Solid (dash-dotted) lines correspond to the GB12
        (SN11~\protect\cite{sn11}) solution.  Thick solid (dashed)
        lines give GZ12 solution (MAID07~\protect\cite{maid},
        which terminates at W = 2~GeV).
        (a) Re[$_nE^{1/2}_{0+}$],
        (b) Im[$_nE^{1/2}_{0+}$],
        (c) Re[$_nM^{1/2}_{1-}$],
        (d) Im[$_nM^{1/2}_{1-}$],
        (e) Re[$_nM^{1/2}_{1+}$], 
        (f) Im[$_nM^{1/2}_{1+}$],
        (g) Re[$_nE^{1/2}_{2-}$], and
        (h) Im[$_nE^{1/2}_{2-}$].
        Vertical arrows indicate mass ($W_R$), and horizontal
        bars show full ($\Gamma$) and partial ($\Gamma_{\pi N}$)
        widths of resonances extracted by the Breit-Wigner fit
        of the $\pi N$ data associated with the SAID solution
        SP06~\protect\cite{piN}.} \label{fig:3}
\end{figure}
\begin{table}[th]
\caption{Breit-Wigner resonance parameters [mass ($W_R$), full
        ($\Gamma$), and partial ($\Gamma_{\pi N}$) widths of
        resonances] associated with the SAID solution SP06
        \protect\cite{piN} obtained from $\pi N$ scattering
        (second column) and neutron helicity amplitudes $A_{1/2}$
        and $A_{3/2}$ (in [(GeV)$^{-1/2}\times 10^{-3}$] units)
        from the GB12 solution (first row), previous SN11
        \protect\cite{sn11} solution (second row), and average
        values from the PDG10~\protect\cite{PDG} (third row).
        \label{tab:tbl1}}
\vspace{2mm}
\begin{tabular}{|c|c|c|c|}
\hline
Resonance       & $\pi N$ SAID               &   $A_{1/2}$  & $A_{3/2}$    \\
\hline
$N(1535)1/2^-$  & $W_{R}$=1547~MeV           &  $-$58$\pm$6 &   \\
                & $\Gamma$=188~MeV           &  $-$60$\pm$3 &   \\
                & $\Gamma _{\pi N}/\Gamma$=0.36&  $-$46$\pm$27& \\
\hline
$N(1650)1/2^-$  & $W_{R}$=1635~MeV           &  $-$40$\pm$10&  \\
                & $\Gamma$=115~MeV           &  $-$26$\pm$8 &  \\
                & $\Gamma _{\pi N}/\Gamma$=1.00& $-$15$\pm$21 & \\
\hline
$N(1440)1/2^+$  & $W_{R}$=1485~MeV           &   48$\pm$4   & \\
                & $\Gamma$=284~MeV           &   45$\pm$15  & \\
                & $\Gamma _{\pi N}/\Gamma$=0.79&  40$\pm$10   & \\
\hline
$N(1520)3/2^-$  & $W_{R}$=1515~MeV           & $-$46$\pm$6  & $-$115$\pm$5 \\
                & $\Gamma$=104~MeV           & $-$47$\pm$2  & $-$125$\pm$2 \\
                & $\Gamma _{\pi N}/\Gamma$=0.63&$-$59$\pm$9   & $-$139$\pm$11\\
\hline
$N(1675)5/2^-$  & $W_{R}$=1674~MeV           & $-$58$\pm$2  & $-$80$\pm$5  \\
                & $\Gamma$=147~MeV           & $-$42$\pm$2  & $-$60$\pm$2  \\
                & $\Gamma _{\pi N}/\Gamma$=0.39& $-$43$\pm$12 & $-$58$\pm$13 \\
\hline
$N(1680)5/2^+$  & $W_{R}$=1680~MeV           &26$\pm$4      & $-$29$\pm$2  \\
                & $\Gamma$=128~MeV           &50$\pm$4      & $-$47$\pm$2  \\
                & $\Gamma _{\pi N}/\Gamma$=0.70&29$\pm$10     & $-$33$\pm$9  \\
\hline
\end{tabular}
\end{table}

Resonance couplings were extracted as in Ref.~\cite{sn11}, are
listed in Table~\ref{tab:tbl1} and compared to the previous SN11
determinations and the Particle Data Group (PDG) averages~\cite{PDG}.
Couplings for the $N(1440)1/2^+$, $N(1520)3/2^-$, and $N(1675)5/2^-$ 
are reasonably close to the SN11 estimates. The value of $A_{1/2}$ 
found for the $N(1535)1/2^-$, using the GB12 fit, is very close to 
the SN11 determination. Using the GZ12 fit, however, the result is 
somewhat larger in magnitude ($-85\pm 15$). A similar feature was 
found for the proton couplings, using this form, in Ref.~\cite{mc12}.  
Using this alternate form, a determination of the $N(1650)1/2^-$ 
$A_{1/2}$ was difficult and resulted in a value, lower in magnitude 
by about 50\% .  For this reason, we consider the uncertainty 
associated with this state to be a lower limit only. No value was 
quoted for the $N(1720)3/2^+$ state. As can be seen in Figs.~\ref{fig:3}, 
the two different fit forms GB12 and GZ12, though similar in shape, 
have opposite signs for the imaginary parts of corresponding multipoles
($_n E^{1/2}_{1+}$ and $_n M^{1/2}_{1+}$) in the neighborhood of the
resonance position, and even the sign can not be determined.  This
is in line with the PDG estimates, which also fail to give signs for
the couplings to this state.

Let us \textbf{summarize}: a comprehensive set of differential cross
section at 26 energies for negative-pion photoproduction on the
neutron, via the reaction $\gamma d\to\pi^-pp$, have been determined
with a JLab tagged-photon beam for incident photon energies from 1.05
to 2.7~GeV. To accomplish a state-of-the-art analysis, we included
new FSI corrections using a diagrammatic technique, taking into
account a kinematical cut with momenta less (more) than 200~MeV/$c$
for slow (fast) outgoing protons.

The updated PWAs examined mainly the effect of new CLAS neutron-target
data on the SAID multipoles and resonance parameters. These new data
have been included in a SAID multipole analysis, resulting in new
SAID solutions, GB12 and GZ12. A major accomplishment of this CLAS
experiment is a substantial improvement in the $\pi^-$-photoproduction
database, adding $626$ new differential cross sections, quadrupling
the world database for $\gamma n\to\pi^-p$ above 1~GeV.  Comparison
to earlier SAID fits, and a lower-energy fit from the Mainz group,
shows that the new solutions are much more satisfactory at higher
energies.

On the experimental side, further improvements in the PWAs await more
data, specifically in the region above 1~GeV, where the number of
measurements for this reaction is small.  Of particular importance
in all energy regions is the need for data obtained involving
polarized photons and polarized targets.  Due to the closing of
hadron facilities, new $\pi^-p\to\gamma n$ experiments are not
planned and only $\gamma n\to\pi^-p$ measurements are possible at
electromagnetic facilities using deuterium targets. Our agreement
with existing $\pi^-$ photoproduction measurements leads us to
believe that these photoproduction measurements are reliable despite
the necessity of using a deuterium target. We hope that new CLAS
$\Sigma$-beam asymmetry measurements for $\vec{\gamma}n\to\pi^-p$,
at E$_\gamma$ = 910 up to 2400~MeV will soon~\cite{dar} provide 
further constraints for the neutron multipoles.

Obviously, any meson photoproduction treatment on the ``neutron"
target requires a FSI study. Generally, FSI depends on the full 
set of kinematical variables of the reaction. In our analysis, the 
FSI correction factor depends on the photon energy, meson production 
angle, and is averaged on the rest of variables in the region of 
``quasi-free" process on the neutron.

\begin{acknowledgement}
The authors acknowledge helpful comments and preliminary fits from
R.~A.~Arndt in the early stages of this work.  We acknowledge 
the outstanding efforts of the CLAS Collaboration who made the 
experiment possible. This work was supported in part by the 
U.S. Department of Energy Grants, by the Russian, by the Russian 
Atomic Energy Corporation, the National Science Foundation, and
the Italian Istituto Nazionale di Fisica Nucleare.
\end{acknowledgement}


\end{document}